\documentclass[prl,twocolumn,showpacs,preprintnumbers,amsmath,amssymb]{revtex4}

\usepackage{graphicx}
\usepackage{dcolumn}
\usepackage{bm}

\begin{document}

\title{The Role of Thermal Phase Fluctuations in Underdoped YBa$_{2}$Cu$_{3}$O$_{7-\delta}$ Films}

\author{Yuri Zuev, John A. Skinta, Mun-Seog Kim}
\thanks{Present Address: Division of Electromagnetic Metrology, Korea Research Institute of Standards and Science, P.O. Box 102, Yuseong, Daejeon, 305-600, Republic of Korea, e-mail: msk2003@kriss.re.kr} 
\author{Thomas R. Lemberger}
\affiliation{Department of Physics, Ohio State University, Columbus, OH 43210-1106}
\author{E. Wertz, K. Wu, and Q. Li}
\affiliation{Department of Physics, Pennsylvania State University,
University Park, PA 16802}

\date{\today}

\begin{abstract}

The effect of thermal phase fluctuations (TPF's) on the $ab$-plane penetration depth, $\lambda (T)$, of thin YBa$_{2}$Cu$_{3}$O$_{7-\delta}$ (YBCO) films is found to be much smaller than expected from the paradigm of cuprates as weakly-coupled 2D superconducting layers. A 2D vortex-pair-unbinding transition is observed, but the effective thickness for fluctuations is the film thickness, not a CuO bilayer thickness. In a strongly underdoped YBCO film, $T_C = 34 K$, TPF's suppress $T_C$ by only about 3 K. They cannot be a significant factor in the suppression of $T_C$ and emergence of the pseudogap with underdoping.

\end{abstract}


\maketitle


Understanding the origin of the pseudogap in underdoped cuprates is one of the central issues in cuprate superconductivity. One proposal is that the temperature, $T^*$, where the pseudogap appears is actually the mean-field superconducting transition temperature, $T_{C0}$, and some sort of superconducting fluctuation suppresses the onset of phase coherence down to the measured $T_C$, which vanishes at strong underdoping. It has been proposed that thermal fluctuations in the phase of the superconducting order parameter are an important effect in cuprates~\cite{KivelsonNature, KivelsonPRL, carlson}. There are two good reasons to expect strong thermal phase fluctuation (TPF) effects in underdoped cuprates: first, coupling between CuO$_2$ (bi)layers is so weak that fluctuations should be quasi-two-dimensional; and second, the superfluid density $n_S$ is small. These factors grow with underdoping.

It is important to know whether TPF's alone can account for a significant suppression of $T_C$. While much of our understanding of TPF's comes from studies of arrays of Josephson-coupled grains, ~\cite{ohta, carlson} a significant effort has gone into more realistic models~\cite{KimCarbotte, CurtyBeck, Timm}. Curty and Beck ~\cite{CurtyBeck} find that fluctuations in the amplitude of the order parameter are more important than phase fluctuations. In the present paper, we look for experimental evidence that thermal phase fluctuations, as understood from simulations of arrays of Josephson-coupled grains, play a significant role in suppressing $T_C$ of heavily underdoped cuprates. 

The feature that we identify as the onset of strong TPF's is a rapid increase in downward curvature in $n_S(T)$, as found in numerical simulations of 2D~\cite{ohta} and quasi-2D~\cite{carlson} superconductors. These simulations find that when TPF's are strong enough to suppress $n_S$ about 35\% below its mean-field value, nonlinear effects (vortex-antivortex pairs; vortex loops) come into play, and as $T$ increases further these nonlinearities rapidly suppress $n_S$ to zero. Not surprisingly, nonlinear effects emerge at the vortex-pair-unbinding transition temperature, $T_{2D}^*$, of a single layer. The downward curvature of $n_S(T)$ begins to grow at $T \approx T_{2D}^*$ and diverges as $(T_C-T)^{-4/3}$ at a 3D-XY transition~\cite{schneider,carlson}. We are interested only in finding the onset of strong fluctuation effects, not details of the critical region, which are often obscured by inhomogeneities anyway. The abrupt downturn in $n_S$ should be enhanced in real samples relative to quasi-2D classical calculations~\cite{carlson} because of two effects: first, as discussed below, TPF's are suppressed for $T$ below an effective Debye temperature, so when they "turn on", they turn on more rapidly than found in simulations; and second, to map simulations onto data, the Josephson coupling energy in the simulations must be proportional to the mean-field superfluid density and hence should decrease as $T$ increases.

Quantitatively, the vortex-pair unbinding temperature, $T_{2D}$, of a 2D superconductor, thickness $d$, is predicted by the well-known relation~\cite{kosterlitz}:
\begin{equation}
\label{T2D} \lambda_{\perp}^{-1}(T_{2D})= \frac{8\pi \mu_0}{\Phi_0^2}T_{2D} = \frac{T_{2D}}{9.8\mathrm{mm\ K}},
\end{equation}
where $\lambda_{\perp}^{-1} \equiv d\lambda^{-2} \propto n_Sd$ is the 2D magnetic penetration depth, and $\Phi_0$ is the flux quantum. Note that the superfluid density is proportional to $\lambda_{\perp}^{-1}$. The hypothetical 2D transition temperature, $T_{2D}^*$, for a single CuO$_2$ bilayer in YBCO follows from Eq.~(\ref{T2D}) with $d$ = 1.17 nm, the center-to-center spacing between bilayers. In effect, Eq.~(\ref{T2D}) predicts a transition when the thermal energy, $k_BT$, equals the superconducting condensation energy in a characteristic volume $2 \pi \xi^2 d$. Measurements of $n_S$ in 2D films of conventional $s$-wave superconductors validate this equation (see, e.g., ref.~\cite{turneaure03}).

To locate the onset of TPF effects, we fit a quadratic (constant curvature) to $n_S(T)$ just below $T_{2D}^*$, then find the temperature where $n_S(T)$ drops below the fit. We first show that in a very thin (4 unit cells) optimally-doped YBCO film a rapid downturn in $n_S$ is clearly present, so there is nothing intrinsic to YBCO that precludes the usual vortex-pair unbinding transition. In the 4-unit-cell-thick film, the observed $T_C$  of 70 K is seen to be about 15 K below its mean-field value, $T_{C0} \approx 85 K$. Thicker (8 and 10 unit cells) optimally-doped YBCO films also show downturns in $n_S$, but at temperatures that suggest that the effective thickness for TPF's is the film thickness. A severely underdoped YBCO film, $T_C = 34 \pm 2 K$, also has a rapid downturn in $n_S$ at a temperature well above $T_{2D}^*$. The data show that $T_C$ is suppressed below $T_{C0}$ by only 3 K. 

Our results seem to conflict with experimental support for strong TPF effects in YBCO provided by ob\-ser\-va\-tions of a critical region about 5 K wide in the superfluid density~\cite{kamal02,anlage,kamal01} and other properties of very clean YBCO crys\-tals~\cite{charalambous,pasler}. We argue that there is something anomalous about these results, however, since they are apparently very sensitive to disorder, and are thus not supported by data on slightly less clean YBCO, on YBCO films ~\cite{paget}, or on BSCCO (see below) that is much more anisotropic than YBCO.



This study reports superfluid density measurements on four samples: films A, B, and C were grown epitaxially by pulsed laser deposition (PLD) on NdGaO$_{3}$ substrates, as detailed elsewhere~\cite{kwon}. They are fully oxygenated. Each 1~cm $\times$ 1~cm film consists of buffer layers of semiconducting Pr$_{0.6}$Y$_{0.4}$Ba$_{2}$Cu$_{3}$O$_{7-\delta}$ that are 12 unit cells thick above, and 8 unit cells thick below, the YBCO film. The underlayer lessens the strain of substrate lattice mismatch on the YBCO, while the capping layer protects the \linebreak YBCO film from damage during handling. Films A, B, and C are nominally 4, 8, and 10 unit cells thick; however, the top and bottom YBCO layers may not be perfectly smooth or homogeneous. Our conclusions are insensitive to this uncertainty. Film D was grown by PLD on a SrTiO$_3$ substrate at 760 C, then annealed at 600 C in a low pressure of oxygen so that it would be severely underdoped. There were buffer layers of PrBa$_{2}$Cu$_{3}$O$_{7-\delta}$ above and below the 40-unit-cell thick YBCO film.

We measured $\lambda_{\perp}^{-1}(T) = d_{\mathrm{film}}/\lambda^2$ with a two-coil mutual inductance technique described in detail elsewhere~\cite{turneaure02}. Each film was centered between two coils roughly 2~mm in diameter and 1 mm long.
In a typical measurement, the sample was cooled to 4.2 K, and a current of roughly 100 $\mu$A was driven at 50 kHz through the coil pressed against the back of the substrate. In this geometry, the induced $ac$ electric field in the film was parallel to the plane of the film and had azimuthal symmetry. It was very nearly uniform through the film thickness because film thicknesses were much less than the magnetic penetration depth, $\lambda$. Thus, the conductivities of all layers in the film were in parallel, and the measurement yielded the sheet conductivity of the {\em entire} film: $\sigma (\omega ,T)d_{\mathrm{film}} = \sigma_{1} (\omega ,T)d_{\mathrm{film}} - i\sigma_{2}(\omega ,T)d_{\mathrm{film}}$. As $T$ slowly increased, the voltage induced in the secondary coil, which was pressed against the film, was measured continuously. $\lambda _{\perp} ^{-1}(T)$ was obtained with an accuracy of about 3\% from $\sigma_2d_{\mathrm{film}}$ by using the relation:
\begin{equation}
\label{lambdef} \lambda_{\perp}^{-1}(T) \equiv \mu_{0} \omega \sigma_{2}(T)d_{\mathrm{film}},
\end{equation}
where $\mu_{0}$ is the permeability of vacuum. Uncertainty in $d_{\mathrm{film}}$ enters only in calculating the 3D penetration depth, $\lambda ^{-2}(T)$, or the 2D penetration depth of a single unit-cell layer.


\begin{figure}[t]
\centering
\includegraphics[width=\columnwidth]{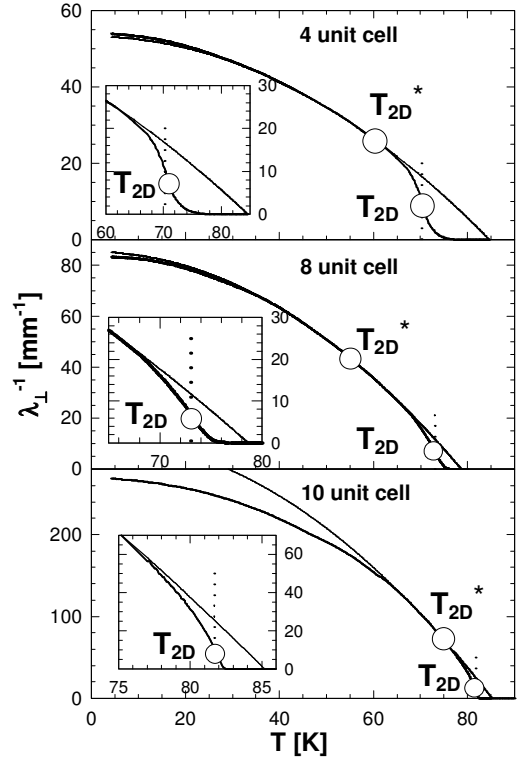}
\caption{$\lambda_{\perp}^{-1} = d_{\mathrm{film}}/\lambda^2$ {\em vs}. $T$ measured at 50 kHz for optimally-doped YBCO films A, B, and C (thick solid
lines). Vertical dotted lines locate the peaks in $\sigma _{1}($50 kHz,$T)$, which very nearly coincide with $T_{2D}$ (open circle). Thin curves are quadratic fits to $\lambda_{\perp}^{-1}(T)$ just below $T_{2D}^*$ (open circle). Insets enlarge the transition regions.}
\end{figure}

Figure 1 shows $\lambda_{\perp}^{-1}(T)$ {\em vs.} $T$ (thick curves) measured for the thin, optimally-doped YBCO films. $\lambda_{\perp}^{-1}(T)$ is quadratic in $T$ at low $T$ presumably due to small disorder and a d-wave superconducting gap. The value of $\lambda_{\perp}^{-1}(0)$ for the 10 unit-cell-thick film is what we routinely observe in thick YBCO films made by pulsed laser deposition. $\lambda_{\perp}^{-1}(0)$ decreases for thicknesses less than 10 unit cells, for reasons yet to be determined.  $T_C$, denoted by vertical dotted lines, is defined as the center of the fluctuation peak in $\sigma_{1}(50$ kHz,$T)$. The two open circles represent $T_{2D}^*$ and $T_{2D}$ calculated from \linebreak Eq.~\ref{T2D} with $d=$ 1 unit cell and $d=$ film thickness, respectively. The thin solid curves are quadratic fits to $n_S(T)$ just below $T_{2D}^*$. They approximate mean-field behavior, $\lambda_{\perp 0}^{-1}(T)$, for $T > T_{2D}^*$.

We expect to see 2D fluctuations in the 4-unit-cell thick film just because it is so thin. Indeed, the top panel of Fig. 1 shows a rapid downturn in $n_S(T) \propto \lambda_{\perp}^{-1}(T)$, highlighted by comparison with the quadratic fit (thin curve). The drop occurs about midway between $T_{2D}^* = 60$ K and $T_{2D} = 70$ K, so we cannot tell whether the effective thickness for fluctuations is one unit cell or the film thickness. Note that the downward curvature in $\lambda_{\perp}^{-1}(T)$ is essentially constant from 4 K to 60 K, meaning that TPF's are negligible below $T_{2D}^* = 60 K$.

We interpret the drop in $\lambda_{\perp}^{-1}(T)$ as a vortex-pair unbinding transition partly because $T_C$ very nearly coincides with $T_{2D}$ obtained from Eq.\ref{T2D}, (circles in Fig. 1), but also because the observed shift of $\sigma_{1}$ and $\lambda_{\perp}^{-1}$ to higher temperature with increasing frequency~\cite{Skintathesis} is similar to behavior seen in $a$-MoGe films~\cite{turneaure03}. Transitions here are slightly broader than in $a$-MoGe films,~\cite{turneaure01} presumably due to slight film inhomogeneity. If the downturn is not due to TPF's, then TPF's are even weaker than we conclude here.
 
In the 8 unit-cell-thick film, the drop in $n_S$ occurs closer to $T_{2D}$ than $T_{2D}^*$, suggesting that the effective thickness for TPF's is the film thickness, not a bilayer thickness. Using the quadratic fit as a proxy for mean-field behavior, we find that TPF's suppress $T_C$ by less than 5 K in this film. The drop in 10 unit-cell-thick film is somewhat broader than in the other films. 

\begin{figure}[t]
\centering
\includegraphics[width=\columnwidth]{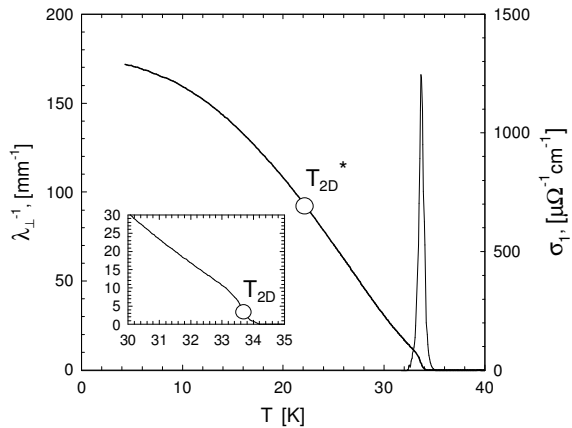} 
\caption{$\lambda_{\perp}^{-1}(T) = d_{\mathrm{film}}/\lambda^2$ and $\sigma _{1}($50 kHz,$T)$ \textit{vs}. $T$ measured at 50 kHz for underdoped YBCO film D (thick curves), 40 unit cells thick. The center of the 2 K wide peak in $\sigma _{1}($50 kHz,$T)$ very nearly coincides with $T_{2D}$ (open circle). Inset shows an enlarged view of the transition region.} 
\end{figure}

We now turn to the main focus of this paper, the role of TPF's in underdoped YBCO. Figure 2 shows $\lambda_{\perp}^{-1}(T)$ and $\sigma_1(T)$ measured at 50 kHz for underdoped film D. From the film's $T_C$ of 34 K, we estimate its oxygen stoichiometry at $O_{6.4}$, so it is strongly underdoped. The peak in $\sigma_1$ is due to critical fluctuations; the peak value of $\sigma_1$ is more than $10^5$ larger than the film's conductivity above $T_C$. The width of the peak, about 2 K, is an upper limit on the inhomogeneity in $T_C$. Nothing in the data at $T \approx T_{2D}^* \approx 22$ K indicates that fluctuations are strong. Instead of arcing downward, $\lambda_{\perp}^{-1}(T)$ develops upward curvature between $T_{2D}^*$ and $T_{2D}$, before finally dropping at $T_{2D} \approx 33.7 K$. Our view is that TPF effects are significant only very close to $T_C$, and that they suppress $T_C$ only a few Kelvins in this strongly underdoped cuprate. This is our central finding.

\begin{figure}[t]
\centering
\includegraphics[width=\columnwidth]{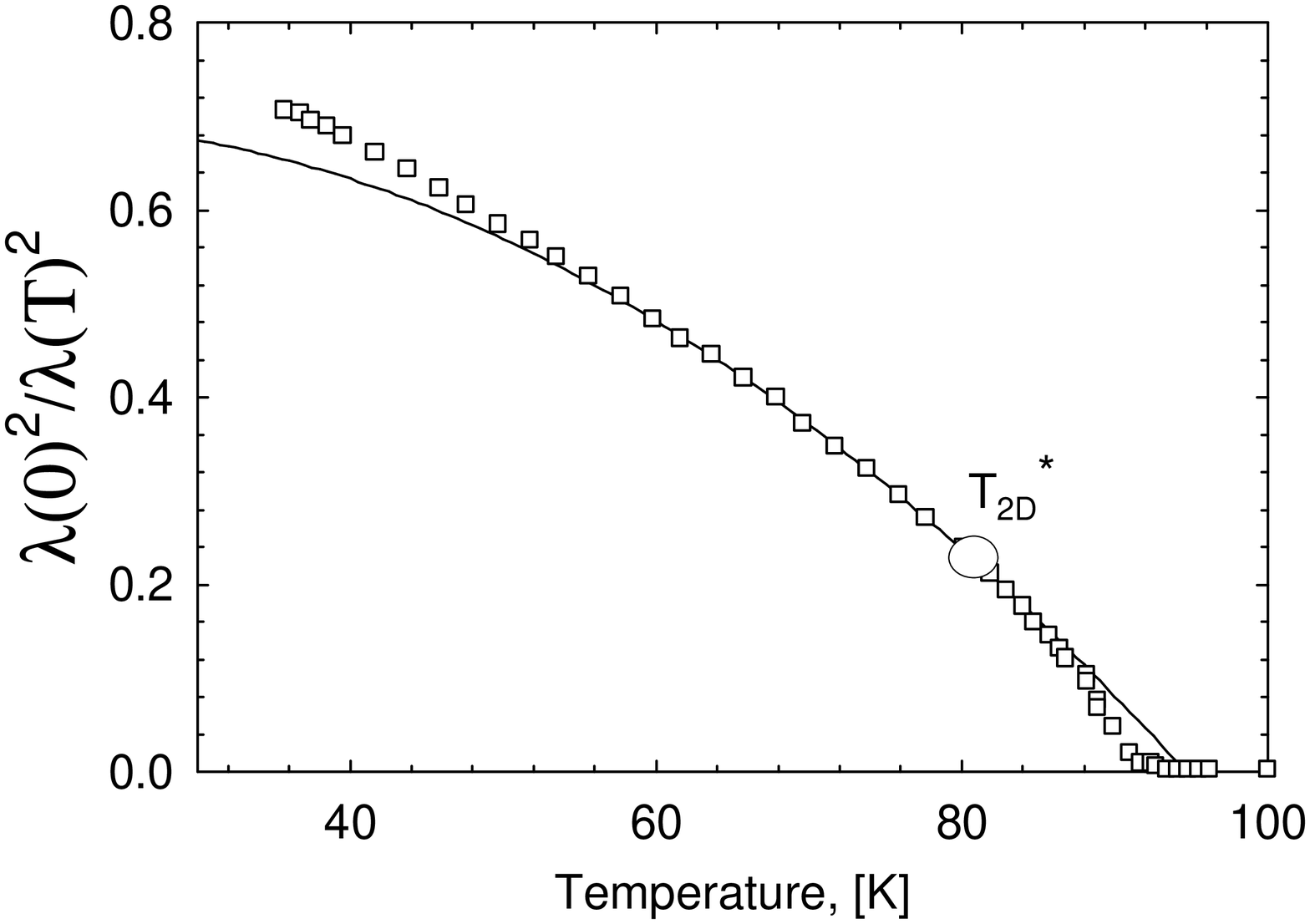} 
\caption{$\lambda^{-2}(T)$ for a single crystal of BSCCO~\cite{lee}. The thin solid curve is a quadratic fit to $\lambda^{-2}(T)$ between 60 K and 76 K. The circle indicates $T_{2D}^* \approx 80 K$.} 
\end{figure}

Since we find TPF's to be weak in YBCO films, while very clean YBCO crystals show strong fluctuations, it is interesting to consider BSCCO, a cuprate in which interlayer coupling is about 100 times weaker than in YBCO. Figure 3 shows measurements~\cite{lee} of $\lambda^{-2}(T)$ measured at 34.7 GHz on a high-quality optimally-doped BSCCO crystal. $\lambda^{-2}(T)$ is linear in $T$ at low $T$, and the low-$T$ slope extrapolates to zero at about 125 K. Lee et al. assigned $T_C = 90 K$ from the peak in $\sigma_1$. The open circle in Fig. 3 marks $T_{2D}^* \approx 80$ K. The thin solid curve is a quadratic fit to data between 60 K and 76 K. Data drop below the fit for $T \ge 88$ K, which we interpret as the onset of strong phase fluctuations. We emphasize that at $88$ K the properly normalized temperature, $T / \lambda^{-2}(T)$ is 2.6 times larger than at $T_{2D}^*$. We argue that TPF's must be weak at 80 K if superconductivity can survive a factor of 2.6 increase in normalized temperature. From the quadratic fit, or from a tangent to the data at 80 K, we estimate that TPF's suppress $T_C$ by about 6 K. It seems that the fluctuations in very clean YBCO crystals are more interesting than simple phase fluctuations.

Why are fluctuations so weak? It is known that TPF's weaken when $T$ drops below an effective "Debye" temperature,~\cite{lemberger, benfatto, turneaure01} but it turns out that this effect cannot account for our results. The effective Debye temperature, $T_Q$, is the temperature where the thermal frequency, $k_BT / \hbar$, equals the characteristic "$R/L$" superfluid relaxation rate, $\Omega(T) \approx \rho_N / \mu _0 \lambda^2(T)$. $\rho_N$ is the normal-state resistivity. We estimate $T_Q \approx$ 45 K, 40 K, and 60 K, for $d$ = 4, 8, and 10 unit cells, respectively, and $T_Q \approx 25 K$ for the underdoped film. Thus, in the optimally-doped films this quantum effect is certainly minor for $T > T_{2D}^*$. For the underdoped film, even though $T_Q$ lies between $T_{2D}^*$ and $T_C$, quantum suppression of TPF's is by an algebraic factor, $\approx 1/[1 + \hbar \Omega(T)/k_BT]$, and would not be strong enough to obscure the dramatic TPF-induced drop in $n_S$.

It seems that electron transport in cuprate superconductors is somehow granular. Grains extend through the film thickness, even for films 40 unit cells thick, hence the effective thickness for phase fluctuations is the film thickness. Granular models have been proposed, e.g., ref.~\cite{Phillips97, MacKenzie, Geshkenbein} describe a model that accounts for many physical properties of cuprates, including the $T$-linear resistivity at optimal doping.

Finally, we comment on fluctuation behavior in YBCO crystals~\cite{kamal01}. In very clean YBCO samples, the 3D critical region extends below $T_{2D}^*$, whereas in the layered superconductor model the 3D critical region must lie above $T_{2D}^*$. For example, Kamal et al.~\cite{kamal01} find that at temperatures from $\sim 85$\% to $\sim 99.9$\% of $T_C \approx 93.78$ K, $\lambda_a^{-2}(T)$ is fitted to within 1\% by the 3D-XY exponent of 2/3: $\lambda_a^{-2}(T)/\lambda_a^{-2}(0) \approx 1.26(1 -T/T_C)^{2/3}$. $T_{2D}^*$ for this crystal is $0.96 T_C$, where $\lambda_a^{-2}(T_{2D}^*)/\lambda_a^{-2}(0) \approx 0.14$. Thus, the paradigm of cuprates as layered superconductors does not find good agreement with fluctuation behavior observed in the superfluid density.


In conclusion, we have made high precision, low frequency  measurements of
$\lambda_{\perp }^{-1}(T)$ and $\sigma_{1}(T)$ in 4 to 40 unit-cell thick YBCO films. Two-dimensional thermal phase fluctuation effects are observed, but the characteristic thickness is the film thickness, not the thickness of a CuO$_2$ bilayer, and TPF's are therefore weaker than expected. Our main point is that even in strongly underdoped YBCO films, thermal phase fluctuations, as they are understood from simulations of Josephson-coupled grains, suppress $T_C$ by only a few Kelvins. They cannot account for the reduction of $T_C$ to zero with underdoping. It may well be that fluctuations of some kind suppress the measured $T_C$ well below its mean-field value, but that fluctuation is subtler than simple thermal phase fluctuations in a quasi-2D superconductor.

The authors gratefully acknowledge useful discussions with C. Jayaprakash, D.G. Stroud, and Y.-B. Kim. This work was supported in part by NSF DMR grant 0203739.








\begin{thebibliography}{88}

\bibitem{KivelsonNature}V. J. Emery, S. A. Kivelson, Nature, {\bf 374}, 434, (1995).
\bibitem{KivelsonPRL}V. J. Emery,  S. A. Kivelson, Phys. Rev. Lett. {\bf 74}, 3253 (1995).
\bibitem{carlson} E.W. Carlson, S.A. Kivelson, V.J. Emery, and E. Manousakis,
Phys. Rev. Lett. {\bf 83}, 612 (1999).
\bibitem{ohta} T. Ohta and D. Jasnow, Phys. Rev. B {\bf 20}, 139
(1979); S. Teitel and C. Jayaprakash, Phys. Rev. Lett. {\bf 51},
1999 (1983); Phys. Rev. B {\bf 27}, 598 (1983); W.Y. Shih and D.
Stroud, {\it ibid.} {\bf 32}, 158 (1985); I.-J. Hwang and D.
Stroud, {\it ibid.} {\bf 57}, 6036 (1998).
\bibitem{KimCarbotte} W. Kim and J.P. Carbotte, J. Low Temp. Phys. {\bf135}, 219 (2004).
\bibitem{CurtyBeck} P. Curty, H. Beck, Phys. Rev. Lett., {\bf 91}, 257002 (2003).
\bibitem{Timm} C. Timm, D. Manske, and K.H. Bennemann, Phys. Rev. B {\bf 66}, 094515 (2002).
\bibitem{turneaure01} S.J. Turneaure, T.R. Lemberger, and J.M.
Graybeal, Phys. Rev. Lett. {\bf 84}, 987 (2000).
\bibitem{lemberger} T.R. Lemberger, A.A. Pesetski, and S.J.
Turneaure, Phys. Rev. B {\bf 61}, 1483 (2000).
\bibitem{benfatto} L. Benfatto, S. Caprara, C. Castellani, A. Paramekanti,
and M. Randeria, Phys. Rev. B 63, 174513 (2001).
\bibitem{kamal02} S. Kamal, D.A. Bonn, N. Goldenfeld, P.J. Hirschfeld, R.
Liang, and W.N. Hardy, Phys. Rev. Lett. {\bf 73}, 1845 (1994).
\bibitem{anlage} S.M. Anlage, J. Mao, J.C. Booth, D.H. Wu, and J.L. Peng,
Phys. Rev. B {\bf 53}, 2792 (1996).
\bibitem{kamal01} S. Kamal, R. Liang, A. Hosseini, D.A. Bonn, and W.N. Hardy,
Phys. Rev. B {\bf 58}, R8933 (1998).
\bibitem{charalambous} M. Charalambous {\em et al.} 
Phys. Rev. Lett. {\bf 83}, 2042 (1999).
\bibitem{pasler} V. Pasler {\em et al.} 
Phys. Rev. Lett. {\bf 81}, 1094 (1998).
\bibitem{schneider} T. Schneider and J.M. Singer, Physica
(Amsterdam) {\bf 341-348C}, 87 (2000).
\bibitem{chakravarty} S. Chakravarty, R.B. Laughlin, D.K. Morr, and C. Nayak,
Phys. Rev. B {\bf 63}, 094503 (2001).
\bibitem{kosterlitz} J.M. Kosterlitz and D.J. Thouless, J.
Phys. C {\bf 6}, 1181 (1973); J.M. Kosterlitz, {\it ibid} {\bf 7},
1046 (1974); V.L. Berezinskii, Sov. Phys. JETP {\bf 32}, 493 (1971).
Phys. Rev. B {\bf 57}, 7986 (1998).
\bibitem{paget} K.M. Paget, B.R. Boyce, and T.R. Lemberger, Phys.
Rev. B {\bf 59}, 6545 (1999).
\bibitem{lee} S.-F. Lee {\em et al.}, 
Phys. Rev. Lett. {\bf 77}, 735 (1996).
\bibitem{turneaure03} S.J. Turneaure, T.R. Lemberger, and J.M.
Graybeal, Phys. Rev. B {\bf 63}, 174505 (2001).
\bibitem{kwon} C. Kwon, Q. Li, X.X. Xi {\it et al}., Appl. Phys. Lett.
{\bf 62}, 1289 (1993); C. Kwon, Q. Li, I. Takeuchi {\it et al}.,
Physica (Amsterdam) {\bf 266C}, 75 (1996).
\bibitem{Phillips97} J.C. Phillips, Proc. Nat'l. Acad. Sci. USA 94, 12771 (1997).
\bibitem{MacKenzie} 
A. Carrington, A. P. Mackenzie and A. Tyler, Phys. Rev. B, {\bf 54}, 3788 (1996)
\bibitem{Geshkenbein} V. B. Geshkenbein, L. B. Ioffe and A. J. Millis, Phys. Rev. Lett., {\bf 80}, 5778 (1998).
\bibitem{turneaure02} S.J. Turneaure, E.R. Ulm, and T.R. Lemberger, J. Appl. Phys. {\bf 79}, 4221 (1996); S.J. Turneaure, A.A. Pesetski, and T.R. Lemberger, {\it ibid.} {\bf 83}, 4334 (1998).
\bibitem{Skintathesis} J.A. Skinta, Ph.D. thesis, Ohio State University, 2000 (unpublished).



\end{thebibliography}
\end{document}